\documentclass[twocolumn,10pt,journal,a4paper]{IEEEtran}
\usepackage[latin1]{inputenc}
\usepackage{amssymb}
\usepackage{latexsym}
\usepackage{mathrsfs}
\usepackage{amsfonts}
\usepackage{amsbsy}
\usepackage{amsmath}
\usepackage{times}
\usepackage{epsfig,epsf,times,amssymb,amsmath}
\usepackage{color}
\usepackage{setspace}
\usepackage{array}

\input epsf
\title{Alternating Optimization Techniques for Power Allocation and Receiver Design in Multihop Wireless Sensor Networks}
\author{Tong Wang, Rodrigo C. de Lamare, \IEEEmembership{Senior Member, IEEE}, and Anke Schmeink, \IEEEmembership{Member, IEEE}
\thanks{Copyright (c) 2013 IEEE. Personal use of this material is permitted. However, permission to use this material for any other purposes must be obtained from the IEEE by sending a request to pubs-permissions@ieee.org.

T. Wang and A. Schmeink are with the Institute for Theoretical Information Technology, RWTH Aachen University, Germany.

R. C. de Lamare is with CETUC / PUC-Rio, Brazil and Department of Electronics, University of York, U.K.

The work was originally funded by the UK MoD, the University of York and was supported by DFG grant SCHM 2643/4-1 and the Alexander von Humboldt Foundation}}

\begin{document}
\maketitle
\thispagestyle{empty}
\pagestyle{empty}
\begin{abstract}
In this paper, we consider a multihop wireless sensor network with multiple relay nodes for each hop where the amplify-and-forward scheme is employed. We present algorithmic strategies to jointly design linear receivers and the power allocation parameters via an alternating optimization approach subject to different power constraints which include global, local and individual ones. Two design criteria are considered: the first one minimizes the mean-square error and the second one maximizes the sum-rate of the wireless sensor network. We derive constrained minimum mean-square error and constrained maximum sum-rate expressions for the linear receivers and the power allocation parameters that contain the optimal complex amplification coefficients for each relay node. An analysis of the computational complexity and the convergence of the algorithms is also presented. Computer simulations show good performance of our proposed methods in terms of bit error rate and sum-rate compared to the method with equal power allocation and an existing power allocation scheme.

\end{abstract}

\section{Introduction}
Recently, wireless sensor networks (WSNs) have attracted a great
deal of research interest because of their unique features that
allow a wide range of applications in the areas of defence,
environment, health and home \cite{Akyildiz}. WSNs are usually
composed of a large number of densely deployed sensing devices which
can transmit their data to the desired destination through multihop
relays \cite{Laneman}. Considering the traditional wireless networks
such as cellular systems, the primary goal in such systems is to
provide high QoS and bandwidth efficiency. The base stations have
easy access to the power supply and the mobile user can replace or
recharge exhausted batteries in the handset \cite{Akyildiz}.
However, power conservation is getting more important, especially
for WSNs. One of the most important constraints on WSNs is the low
power consumption requirement as sensor nodes carry limited,
generally irreplaceable, power sources. Therefore, low complexity
and high energy efficiency are the most important design
characteristics for WSNs. In a cooperative WSN, nodes relay signals
to each other in order to propagate redundant copies of the same
signals to the destination nodes. Among the existing relaying
schemes, the amplify-and-forward (AF) and the decode-and-forward
(DF) are the most popular approaches \cite{Hong,armo}. In the AF
scheme, the relay nodes amplify the received signal and rebroadcast
the amplified signals toward the destination nodes. In the DF
scheme, the relay nodes first decode the received signals and then
regenerate new signals to the destination nodes subsequently.

Some power allocation methods have been proposed for WSNs to obtain
the best possible signal-to-noise ratio (SNR) or best possible
quality of service (QoS) \cite{Quek,Vardhe} at the destinations.  By
adjusting appropriately the power levels used for the links between
the sources, the relays and the destinations, significant
performance gains can be obtained for a given power budget. Most of
the research on power allocation for WSNs are based on the
assumption of perfect synchronization and available channel state
information (CSI) at each node. A WSN is said to have full CSI when
all of its nodes have access to accurate and up-to-date CSI. When
full CSI is available to all the nodes, the power of each node can
be optimally allocated to improve the system efficiency and lower
the outage probability \cite{Li} or bit error rate (BER)
\cite{Rodrigo,jpais}.

In WSNs, some power allocation problems can be formulated as
centralized or distributed optimization problems subject to power
constraints on certain groups of signals. For the centralized
schemes \cite{Dohler,Adeane}, a network controller is required which
is responsible for monitoring the information of the whole network
such as the CSI and SNR, calculating the optimum power allocation
parameters of each link and sending them to all nodes via feedback
channels. This scheme considers all the available links but it has
two major drawbacks. The first one is the high computational burden
and storage demand at the network controller. The second one is that
it requires a significant amount of control information provided by
feedback channels which leads to a loss in bandwidth efficiency. For
the distributed schemes \cite{Chen}, each node only needs to have
the knowledge of its 'partner' information and calculate its own
power allocation parameter. Therefore, a distributed scheme requires
less control information and is ideally suited to WSNs. However, the
performance of distributed schemes is inferior to centralized
schemes \cite{tds}.

Due to the inherent limitations in the sensor node size, power and
cost \cite{Akyildiz}, they are only able to communicate in a short
range. Therefore, multihop communication \cite{Boyer} is employed to
enhance the coverage of WSNs. By using multihop transmissions, the
rapid decay of the received signal which is caused by the increased
transmission distance can be overcome. Moreover, pathways around the
obstacles between the source and destination can be provided to
avoid the signal shadowing \cite{Gharavi}. Several works about power
allocation of multihop transmission systems have been proposed in
\cite{Shelby}-\cite{Hasna}. The work reported in \cite{Shelby}
develops a cross-layer model for multihop communication and analyzes
the energy consumption of multihop topologies with equal distance
and optimal node spacing. Centralized and Distributed schemes for
power allocation are presented to minimize the total transmission
power under a constraint on the BER at the destination in
\cite{Maham} and \cite{Lau}. In \cite{Farhadi}, two optimal power
allocation schemes are proposed to maximize the instantaneous
received SNR under short-term and long-term power constraints. In
\cite{Hasna}, the outage probability is considered as the
optimization criterion to derive the optimal power allocation
schemes under a given power budget for both regenerative and
non-regenerative systems.

In this paper, we consider a general multihop WSN where the AF
relaying scheme is employed. The proposed strategy is to jointly
design the linear receivers and the power allocation parameters that
contain the optimal complex amplification coefficients for each
relay node via an alternating optimization approach
\cite{Csiszar,Niesen}. Two kinds of linear receivers are designed,
the minimum mean-square error (MMSE) receiver and the maximum
sum-rate (MSR) receiver. They can be considered as solutions to
constrained optimization problems where the objective function is
the mean-square error (MSE) cost function or the sum-rate (SR) and
the constraint is a bound on the power levels among the relay nodes.
Then, the constrained MMSE or MSR expressions for the linear
receiver and the power allocation parameter can be derived. The
major novelty in these strategies presented here is that they are
applicable to general multihop WSNs with multi source nodes and
destination nodes, as opposed to the simple two-hop WSNs with one
pair of source-destination nodes \cite{Quek,[16],Huang}. Unlike the
previous works on the power allocation for multihop systems in
\cite{Shelby}-\cite{Hasna}, in our work, the power allocation and
receiver coefficients are jointly optimized. The joint strategies
were proposed for a two-hop WSN with multiple relay nodes in our
previous work \cite{Wang12}. In order to increase the applicability
of our investigation, in this paper, we develop joint strategies for
general multihop WSNs. They can be considered as an extension of the
strategies proposed for the two-hop WSNs and more complex
mathematical derivations are presented. Moreover, different kinds of
power constraints can be considered and compared. For the MMSE
receiver, we present three strategies where the allocation of power
level across the relay nodes is subject to global, local and
individual power constraints. Another fundamental contribution of
this work is the derivation of a closed-form solution for the
Lagrangian multiplier ($\lambda$) that arises in the expressions of
the power allocation parameters. For the MSR receiver, the local
power constraints are considered. We propose a strategy that employs
iterations with the Generalized Rayleigh Quotient \cite{Juday} to
solve the optimization problem in an alternating fashion. Some
preliminary results of these work have been reported in
\cite{Wang_ICASSP}. The main contributions of this paper can be
summarized as:
\begin{description}
\item[1)]Constrained MMSE expressions for the design of linear receivers and power allocation parameters for multihop WSNs. The constraints include the global, local and individual power constraints.

\item[2)]Constrained MSR expressions for the design of linear receivers and power allocation parameters for multihop WSNs subject to local power constraints.

\item[3)]Alternating optimization algorithms that compute the linear receivers and power allocation parameters in 1) and 2) to minimize the mean-square error or maximize the sum-rate of the WSN.

\item[4)]Analysis of the computational complexity and the convergence of the proposed optimization algorithms.
\end{description}

The rest of this paper is organized as follows.  Section II describes the general multihop WSN system model. Section III develops three joint MMSE receiver design and power allocation strategies subject to three different power constraints. Section IV develops the joint MSR receiver design and power allocation strategy subject to local power constraints. Section V contains an analysis of the computational complexity and the convergence. Section VI presents and discusses the simulation results, while Section VII provides some concluding remarks.

\section{Multihop WSN System Model}
Consider a general $m$-hop wireless sensor network (WSN) with
multiple parallel relay nodes for each hop, as shown in Fig.
1.  The WSN consists of $N_0$ source nodes, $N_m$ destination nodes and
$N_r$ relay nodes which are separated into $m-1$ groups:
$N_1$, $N_2$, ... , $N_{m-1}$.
The index refers to the number of nodes after a given phase of transmission starting with $0$ and going up to $m-1$. The proposed optimization algorithms in this paper refer to a particular instance, for which the roles of the nodes acted as sources, relays and destinations have been pre-detemined. In subsequent time slots these roles can be swapped so that all nodes can actually work as potential sources.
We concentrate
on a time division scheme with perfect synchronization, for which
all signals are transmitted and received in separate time slots.
The sources first broadcast the $N_0\times1$ signal vector
\textbf{s} which contains $N_0$ signals in parallel to the first group of relay nodes. We
consider an amplify-and-forward (AF) cooperation protocol in this
paper. An extension to other cooperation protocols is straightforward. Each group of relay nodes receives the signals, amplifies and rebroadcasts them to
the next group of relay nodes (or the destination nodes). In practice, we
need to consider the constraints on the transmission policy. For
example, each transmitting node would transmit during only one phase. In our
WSN system, we assume that each group of relay nodes transmits the
signal to the nearest group of relay nodes (or the destination nodes)
directly. We can use a block diagram to indicate the multihop
WSN system as shown in Fig. 2.
\begin{figure}[!htb]
\begin{center}
\hspace*{0em}{\includegraphics[width=8cm, height=5cm]{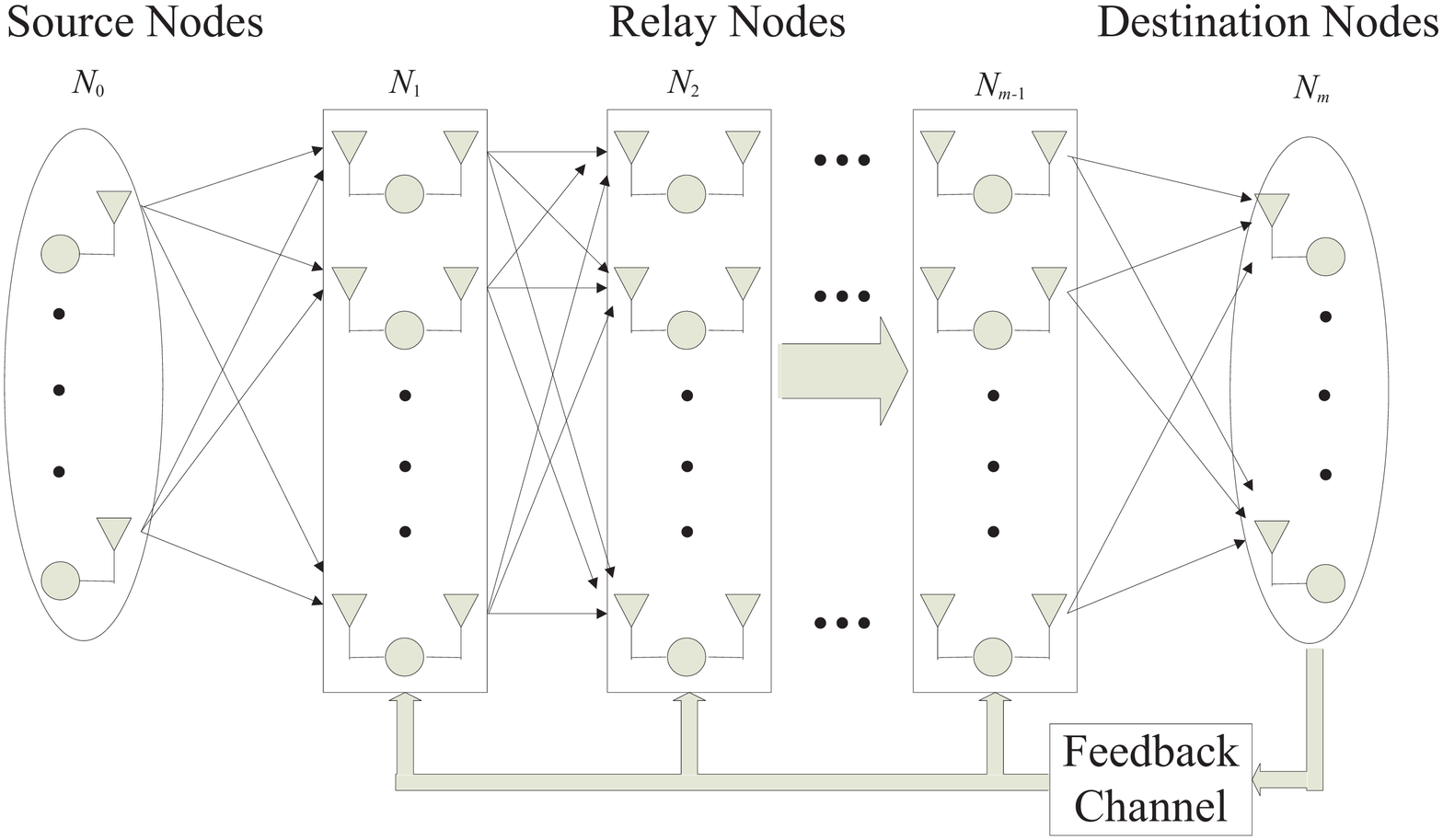}}
\vspace*{0.0em} \caption{An $m$-hop WSN with $N_0$
source nodes, $N_m$ destination nodes and $N_r$ relay nodes.}
\end{center}
\end{figure}
\begin{figure}[!htb]
\begin{center}
\hspace*{0em}{\includegraphics[width=8cm, height=3cm]{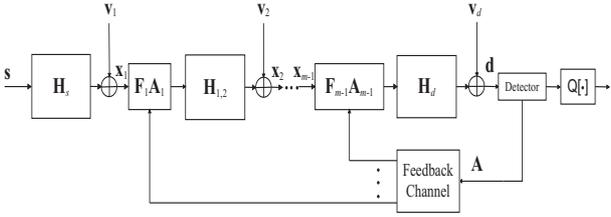}}
\vspace*{0.0em} \caption{Block diagram of the multihop WSN
system.}
\end{center}
\end{figure}
Let $\textbf{H}_s$ denote the  $N_1\times{N_0}$
channel matrix between the source nodes and the first group of relay nodes,
$\textbf{H}_d$ denote the $N_m\times{N_{m-1}}$ channel
matrix between the $(m-1)$th group of relay nodes and destination nodes, and
$\textbf{H}_{i-1,i}$ denote the
$N_{i}\times{N_{i-1}}$ channel matrix between two groups of
relay nodes as described by
\begin{equation}
 \textbf{H}_s=
 \begin{bmatrix}
 \textbf{h}_{s,1}\\
 \textbf{h}_{s,2}\\
 \vdots\\
 \textbf{h}_{s,N_{1}}\\
 \end{bmatrix},~~~
 \textbf{H}_{d}=
 \begin{bmatrix}
 \textbf{h}_{m-1,1}\\
 \textbf{h}_{m-1,2}\\
 \vdots\\
 \textbf{h}_{m-1,N_{m}}\\
 \end{bmatrix},~~~
 \textbf{H}_{i-1,i}=
 \begin{bmatrix}
 \textbf{h}_{i-1,1}\\
 \textbf{h}_{i-1,2}\\
 \vdots\\
 \textbf{h}_{i-1,N_{i}}\\
 \end{bmatrix},~~~
\end{equation}
where $\textbf{h}_{s,j}=[h_{s,j,1},h_{s,j,2},...,h_{s,j,N_0}]$ for $j=1,2,...,N_1$ denote the channel coefficients between the source nodes and the $j$th relay of the first group of relay nodes, $\textbf{h}_{m-1,j}=[h_{m-1,j,1},h_{m-1,j,2},...,h_{m-1,j,N_{m-1}}]$ for $j=1,2,...,N_m$ denote the channel coefficients between the $(m-1)$th group of relay nodes and the $j$th destination node. Further, $\textbf{h}_{i-1,j}=[h_{i-1,j,1},h_{i-1,j,2},...,h_{i-1,j,N_{i-1}}]$ for $j=1,2,...,N_i$ denotes the channel coefficients between the $(i-1)$th group of relay nodes and the $j$th relay of the $i$th group of relay nodes. The received signal at the $i$th group of relay nodes ($\textbf{x}_i$) for each phase can be expressed as:

Phase 1:
\begin{equation}
\label{eq:mhop:MSR:(2)}
\textbf{x}_1=\textbf{H}_{s}\textbf{s}+\textbf{v}_{1},
\end{equation}
\begin{equation}
\textbf{y}_1=\textbf{F}_1\textbf{x}_1,
\end{equation}

Phase 2:
\begin{equation}
\textbf{x}_2=\textbf{H}_{1,2}\textbf{A}_1\textbf{y}_1+\textbf{v}_{2},
\end{equation}
\begin{equation}
\textbf{y}_2=\textbf{F}_2\textbf{x}_2,
\end{equation}
\vdots

Phase $i$: ($i=3, 4, ... , m-1$)
\begin{equation}
\textbf{x}_i=\textbf{H}_{i-1,i}\textbf{A}_{i-1}\textbf{y}_{i-1}+\textbf{v}_{i},
\end{equation}
\begin{equation}
\label{eq:mhop:MSR:(7)}
\textbf{y}_i=\textbf{F}_i\textbf{x}_i,
\end{equation}

At the destination nodes, the received signal can be expressed as
\begin{equation}
\label{eq:mhop:MSR:(8)}
\textbf{d}=\textbf{H}_{d}\textbf{A}_{m-1}\textbf{y}_{m-1}+\textbf{v}_d,
\end{equation}
where \textbf{v} is a zero-mean circularly symmetric complex
additive white Gaussian noise (AWGN) vector with covariance matrix
$\sigma^2\textbf{I}$. The matrix $\textbf{A}_i={\rm diag}\{a_{i,1},a_{i,2},...,a_{i,N_i}\}$ is a diagonal matrix whose
elements represent the amplification coefficient of each relay of
the $i$th group. The matrix $\textbf{F}_i={\rm diag}\{E(|x_{i,1}|^2),E(|x_{i,2}|^2),$ $...,E(|x_{i,N_i}|^2)\}^{-\frac{1}{2}}$ denotes the normalization matrix which can normalize the power of the received signal for each relay of the $i$th group of relays. Please note that the property of the matrix vector multiplication $\textbf{Ay}=\textbf{Ya}$ will be used in the next section, where \textbf{Y} is the diagonal matrix form of the vector \textbf{y} and \textbf{a} is the vector form of the diagonal matrix \textbf{A}. At the receiver, a linear detector is considered where the optimal filter and optimal amplification coefficients are calculated. The optimal amplification coefficients are transmitted to the relays through the feedback channel. The block marked with a Q[$\cdot$] represents a decision device. In our proposed designs, the full CSI of the system is assumed to be known at all the destination nodes. In practice, a fusion center \cite{Varshney} which contains the destination nodes is responsible for gathering the CSI, computing the optimal linear filters and the optimal amplification coefficients. The fusion center also recovers the transmitted signal of the source nodes and transmits the optimal amplification coefficients to the relay nodes via a feedback channel.

\section{Proposed Joint MMSE Design of the Receiver and Power Allocation}
In this section, three constrained optimization problems are
proposed to describe the joint design of the linear receiver
(\textbf{W}) and the power allocation parameter (\textbf{a}) subject
to a global, local and individual power constraints. Extensions to
nonlinear receiver and advanced parameter estimation techniques are
also possible \cite{spa}-\cite{mdfpic}. They impose different power
limitations on all the relay nodes, each group of relay nodes and
each relay node, respectively. The assumptions of these power
constraints could determine the degrees of freedom for allocating
the power among the relay nodes which will affect the performance
and the lifetime of the networks.
\subsection{MMSE Design with a Global Power Constraint}
We first consider the case where the total power of all the relay nodes is limited to $P_T$. The proposed design can be formulated as the following optimization problem
\begin{equation}
\begin{split}
[\textbf{W}_{\rm opt},\textbf{a}_{1,\rm opt},...,\textbf{a}_{m-1,\rm opt}]& ~=\arg\min_{\textbf{W},\textbf{a}_1,...,\textbf{a}_{m-1}}E[\|\textbf{s}-\textbf{W}^H\textbf{d}\|^2],  \\
{~~~~~~~\textrm {subject to}} & ~ \sum_{i=1}^{m-1}P_i=P_T
\end{split}
\end{equation}
where $(\cdot)^H$ denotes the complex-conjugate (Hermitian) transpose, $P_i$ is the transmitted power of the $i$th group of relay nodes, and $P_i=N_{i+1}\textbf{a}_i^H\textbf{a}_i$. In (9), we employ the equality as the constraint other than the inequality because it is obvious that the more available total power can be used the better performance can be achieved, which means that even if we set the constraint to be an inequality, the best performance will be achieve when the power is set to be the maximum bound (i.e. $P_T$). In addition, for such optimization problems it is easier to derive algorithmic solutions for equality constraints rather than inequality constraints.

To solve this constrained optimization problem, we adopt an alternating optimization approach whose global convergence has been established in \cite{Csiszar} and \cite{Niesen} so that the global minimum value can be achieved. We modify the MSE cost function using the method of Lagrange multipliers \cite{Haykin} which yields the following Lagrangian function
\begin{equation}
\begin{split}
\mathcal{L}=&E[\|\textbf{s}-\textbf{W}^H\textbf{d}\|^2] + \lambda(\sum_{i=1}^{m-1}N_{i+1}\textbf{a}_i^H\textbf{a}_i-P_T) \\
           =&E(\textbf{s}^H\textbf{s})-E(\textbf{d}^H\textbf{Ws})-E(\textbf{s}^H\textbf{W}^H\textbf{d})+E(\textbf{d}^H\textbf{WW}^H\textbf{d})\\
            &+ \lambda(\sum_{i=1}^{m-1}N_{i+1}\textbf{a}_i^H\textbf{a}_i-P_T).
\end{split}
\label{eq:mhop:MMSE:(1)}
\end{equation}
By fixing $\textbf{a}_1,...,\textbf{a}_{m-1}$ and setting the gradient of $\mathcal{L}$ in (\ref{eq:mhop:MMSE:(1)}) with respect to the conjugate of the filter $\textbf{W}$ equal to zero, where $(\cdot)^*$ denotes the complex-conjugate, we get
\begin{equation}
\begin{split}
\textbf{W}_{\rm opt}=&[E(\textbf{dd}^H)]^{-1}E(\textbf{ds}^H)\\
=&\left[\textbf{H}_d\textbf{A}_{m-1}E(\textbf{y}_{m-1}\textbf{y}_{m-1}^H)\textbf{A}_{m-1}^H\textbf{H}_d^H+\sigma_n^2\textbf{I}\right]^{-1}\\
&\times\textbf{H}_d\textbf{A}_{m-1}E(\textbf{y}_{m-1}\textbf{s}^H).
\end{split}
\label{eq:mhop:MMSE:(2)}
\end{equation}
The optimal expression for $\textbf{a}_{m-1}$ is obtained by equating the partial derivative of $\mathcal{L}$ with respect to $\textbf{a}_{m-1}^*$ to zero
\begin{equation}
\begin{split}
\frac{\partial \mathcal{L}}{\partial \textbf{a}_{m-1}^*}=&-E(\frac{\partial \textbf{d}^H}{\partial \textbf{a}_{m-1}^*}\textbf{Ws})+E(\frac{\partial \textbf{d}^H}{\partial \textbf{a}_{m-1}^*}\textbf{WW}^H\textbf{d})\\
&+N_m\lambda \textbf{a}_{m-1}\\
=&-E(\textbf{Y}_{m-1}^H\textbf{H}_d^H\textbf{Ws})\\
 &+E[\textbf{Y}_{m-1}^H\textbf{H}_d^H\textbf{WW}^H(\textbf{H}_d\textbf{Y}_{m-1}\textbf{a}_{m-1}+\textbf{v}_d)]\\
 &+N_m\lambda \textbf{a}_{m-1}\\
=&\textbf{0}.
\end{split}
\end{equation}
Therefore, we obtain
\begin{equation}
\begin{split}
\textbf{a}_{m-1,\rm opt}=&[E(\textbf{Y}_{m-1}^H\textbf{H}_d^H\textbf{WW}^H\textbf{H}_d\textbf{Y}_{m-1})+N_m\lambda \textbf{I}]^{-1}\\
&\times E(\textbf{Y}_{m-1}^H\textbf{H}_d^H\textbf{Ws})\\
=&[\textbf{H}_d^H\textbf{WW}^H\textbf{H}_d\circ E(\textbf{y}_{m-1}\textbf{y}_{m-1}^H)^*+N_m\lambda \textbf{I}]^{-1}\\
&\times[\textbf{H}_d^H\textbf{W}\circ E(\textbf{y}_{m-1}\textbf{s}^H)^*\textbf{u}]
\end{split}
\label{eq:mhop:MMSE:(3)}
\end{equation}
where $\circ$ denotes the Hadamard (element-wise) product and $\textbf{u}=[1,1,...,1]^T$.

Similarly, for $i=2,3,...m-1$, we have
\begin{equation}
\begin{split}
\frac{\partial \mathcal{L}}{\partial \textbf{a}_{i-1}^*}&=-E(\frac{\partial \textbf{d}^H}{\partial \textbf{a}_{i-1}^*}\textbf{Ws})+(\frac{\partial \textbf{d}^H}{\partial \textbf{a}_{i-1}^*}\textbf{WW}^H\textbf{d})+N_i\lambda \textbf{a}_{i-1}\\
&=\textbf{0}
\end{split}
\end{equation}
where
\begin{equation}
\frac{\partial \textbf{d}^H}{\partial \textbf{a}_{i-1}^*}=\textbf{Y}_{i-1}^H\left(\prod_{k=i}^{m-1}\textbf{H}_{k-1,k}^H\textbf{F}_k^H\textbf{A}_{k}^H\right)\textbf{H}_d^H.
\end{equation}
Let
\begin{equation}
\textbf{B}_{i-1}=\prod_{k=i}^{m-1}\textbf{H}_{k-1,k}^H\textbf{F}_k^H\textbf{A}_{k}^H.
\end{equation}
Then, we get
\begin{equation}
\begin{split}
\textbf{a}_{i-1,\rm opt}=&[E(\textbf{Y}_{i-1}^H\textbf{B}_{i-1}\textbf{H}_d^H\textbf{WW}^H\textbf{H}_d\textbf{B}_{i-1}^H\textbf{Y}_{i-1})+N_i\lambda \textbf{I}]^{-1}\\
&\times E(\textbf{Y}_{i-1}^H\textbf{B}_{i-1}\textbf{H}_d^H\textbf{W}\textbf{s})\\
=&[\textbf{B}_{i-1}\textbf{H}_d^H\textbf{WW}^H\textbf{H}_d\textbf{B}_{i-1}^H\circ E(\textbf{y}_{i-1}\textbf{y}_{i-1}^H)^*+N_i\lambda \textbf{I}]^{-1}\\
&\times[\textbf{B}_{i-1}\textbf{H}_d^H\textbf{W}\circ E(\textbf{y}_{i-1}\textbf{s}^H)^*\textbf{u}].
\end{split}
\label{eq:mhop:MMSE:(4)}
\end{equation}
From (\ref{eq:mhop:MMSE:(3)}) and (\ref{eq:mhop:MMSE:(4)}), we conclude that
\begin{equation}
\begin{split}
\textbf{a}_{i,\rm opt}=&[E(\textbf{Y}_{i}^H\textbf{B}_{i}\textbf{H}_d^H\textbf{WW}^H\textbf{H}_d\textbf{B}_{i}^H\textbf{Y}_{i})+N_{i+1}\lambda \textbf{I}]^{-1}\\
&\times E(\textbf{Y}_{i}^H\textbf{B}_{i}\textbf{H}_d^H\textbf{W}\textbf{s})\\
=&[\textbf{B}_{i}\textbf{H}_d^H\textbf{WW}^H\textbf{H}_d\textbf{B}_{i}^H\circ E(\textbf{y}_{i}\textbf{y}_{i}^H)^*+N_{i+1}\lambda \textbf{I}]^{-1}\\
&\times[\textbf{B}_{i}\textbf{H}_d^H\textbf{W}\circ E(\textbf{y}_{i}\textbf{s}^H)^*\textbf{u}]
\end{split}
\label{eq:mhop:MMSE:(5)}
\end{equation}
where
\begin{equation}
\textbf{B}_{i} = \left\{ \begin{array}{ll}
\prod_{k=i+1}^{m-1}\textbf{H}_{k-1,k}^H\textbf{F}_k^H\textbf{A}_{k}^H, & \textrm{for $i=1,2,...,m-2$,}\\\textbf{I}, & \textrm{for $i=m-1$}.\\
\end{array}\right.
\end{equation}
Please see the Appendix to find the expressions of $\textbf{F}_i$, $E(\textbf{y}_{i}\textbf{y}_{i}^H)$, and $E(\textbf{y}_{i}\textbf{s}^H)$. The expressions in (\ref{eq:mhop:MMSE:(2)}) and (\ref{eq:mhop:MMSE:(5)}) depend on each other. Thus, it is necessary to iterate them with an initial value of $\textbf{a}_i$ ($i=1,2,...,m-1$) to obtain the solutions.
\\
\indent The Lagrange multiplier $\lambda$ can be determined by solving
\begin{equation}
\label{eq:mhop:MMSE:(6)}
\sum_{i=1}^{m-1}N_{i+1}\textbf{a}_{i,\rm opt}^H\textbf{a}_{i,\rm opt}=P_T.
\end{equation}
Let
\begin{equation}
\label{eq:mhop:MMSE:(7)}
\boldsymbol{\phi}_i=E(\textbf{Y}_{i}^H\textbf{B}_{i}\textbf{H}_d^H\textbf{WW}^H\textbf{H}_d\textbf{B}_{i}^H\textbf{Y}_{i})
\end{equation}
and
\begin{equation}
\label{eq:mhop:MMSE:(8)}
\textbf{z}_i=E(\textbf{Y}_{i}^H\textbf{B}_{i}\textbf{H}_d^H\textbf{W}\textbf{s}).
\end{equation}
Then, we get
\begin{equation}
\textbf{a}_i=(\boldsymbol{\phi}_i+N_{i+1}\lambda \textbf{I})^{-1}\textbf{z}_i.
\end{equation}
When $\lambda$ is a real value, we have
\begin{equation}
[(\boldsymbol{\phi}_i+N_{i+1}\lambda \textbf{I})^{-1}]^H = [(\boldsymbol{\phi}_i+N_{i+1}\lambda \textbf{I})^{H}]^{-1}=(\boldsymbol{\phi}_i+N_{i+1}\lambda \textbf{I})^{-1}.
\end{equation}
Equation (\ref{eq:mhop:MMSE:(6)}) becomes
\begin{equation}
\label{eq:mhop:MMSE:(9)}
\sum_{i=1}^{m-1}N_{i+1}\textbf{z}_i^H(\boldsymbol{\phi}_i+N_{i+1}\lambda \textbf{I})^{-1}(\boldsymbol{\phi}_i+N_{i+1}\lambda \textbf{I})^{-1}\textbf{z}_i=P_T.
\end{equation}
Using an eigenvalue decomposition (EVD), we have
\begin{equation}
\boldsymbol{\phi}_i=\textbf{Q}_i\boldsymbol{\Lambda}_i\textbf{Q}_i^{-1}
\end{equation}
where $\boldsymbol{\Lambda}_i={\rm diag}\{\alpha_{i,1},\alpha_{i,2},...,\alpha_{i,M_i},0,...,0\}$ consists of eigenvalues of $\boldsymbol{\phi}_i$ and $M_i={\rm min}\{N_0,N_i,N_m\}$. Then, we get
\begin{equation}
\boldsymbol{\phi}_i+N_{i+1}\lambda \textbf{I}=\textbf{Q}_i(\boldsymbol{\Lambda}_i+N_{i+1}\lambda\textbf{I})\textbf{Q}_i^{-1}.
\end{equation}
Therefore, (\ref{eq:mhop:MMSE:(9)}) can be expressed as
\begin{equation}
\label{eq:mhop:MMSE:(10)}
\sum_{i=1}^{m-1}N_{i+1}\textbf{z}_i^H\textbf{Q}_i(\boldsymbol{\Lambda}_i+N_{i+1}\lambda\textbf{I})^{-2}\textbf{Q}_i^{-1}\textbf{z}_i=P_T.
\end{equation}
Using the properties of the trace operation, (\ref{eq:mhop:MMSE:(10)}) can be written as
\begin{equation}
\label{eq:mhop:MMSE:(11)}
\sum_{i=1}^{m-1}N_{i+1}tr\left((\boldsymbol{\Lambda}_i+N_{i+1}\lambda\textbf{I})^{-2}\textbf{Q}_i^{-1}\textbf{z}_i\textbf{z}_i^H\textbf{Q}_i\right)=P_T.
\end{equation}
Defining $\textbf{C}_i=\textbf{Q}_i^{-1}\textbf{z}_i\textbf{z}_i^H\textbf{Q}_i$, (11) becomes
\begin{equation}
\sum_{i=1}^{m-1}\sum_{j=1}^{N_i}N_{i+1}(\alpha_{i,j}+N_{i+1}\lambda)^{-2}\textbf{C}_i(j,j)=P_T.
\end{equation}
Since $\boldsymbol{\phi}_i$ is a matrix with at most rank $M_i$, only the first $M_i$ columns of $\textbf{Q}_i$ span the column space of $E(\textbf{Y}_{i}^H\textbf{B}_{i}\textbf{H}_d^H\textbf{W}\textbf{s})^H$ which causes the last $(N_i-M_i)$ columns of $\textbf{z}_i^H\textbf{Q}_i$ to become zero vectors and the last $(N_i-M_i)$ diagonal elements of $\textbf{C}_i$ are zero. Therefore, we obtain the $\{\sum_{i=1}^{m-1}2M_i\}$th-order polynomial in $\lambda$
\begin{equation}
\sum_{i=1}^{m-1}\sum_{j=1}^{M_i}N_{i+1}(\alpha_{i,j}+N_{i+1}\lambda)^{-2}\textbf{C}_i(j,j)=P_T.
\end{equation}

\subsection{MMSE Design with Local Power Constraints}
Secondly, we consider the case where the total power of the relay nodes in each group is limited to some value $P_{T,i}$. The proposed method can be considered as the following optimization problem
\begin{equation}
\begin{split}
&[\textbf{W}_{\rm opt},\textbf{a}_{1,\rm opt},...,\textbf{a}_{m-1,\rm opt}] =\arg\min_{\textbf{W},\textbf{a}_1,...,\textbf{a}_{m-1}}E[\|\textbf{s}-\textbf{W}^H\textbf{d}\|^2], \\
&~~~~~~~~~{\textrm {subject to}}  ~ P_i=P_{T,i},~i=1,2,...,m-1,
\end{split}
\end{equation}
where $P_i$ as defined above is the transmitted power of the $i$th group of relays, and $P_i=N_{i+1}\textbf{a}_i^H\textbf{a}_i$.
Using the method of Lagrange multipliers again, we obtain the following Lagrangian function
\begin{equation}
\mathcal{L}=E[\|\textbf{s}-\textbf{W}^H\textbf{d}\|^2] + \sum_{i=1}^{m-1}\lambda_i(N_{i+1}\textbf{a}_i^H\textbf{a}_i-P_{T,i}). \\
\end{equation}
Following the same steps as described in Section III.A, we get the same optimal expression for $\textbf{W}$ as in (\ref{eq:mhop:MMSE:(2)}). The optimal expression for the power allocation vector $\textbf{a}_i$ is different from (\ref{eq:mhop:MMSE:(5)}) and is given by
\begin{equation}
\begin{split}
\textbf{a}_{i,\rm opt}=&[\textbf{B}_{i}\textbf{H}_d^H\textbf{WW}^H\textbf{H}_d\textbf{B}_{i}^H\circ E(\textbf{y}_{i}\textbf{y}_{i}^H)^*+N_{i+1}\lambda_i \textbf{I}]^{-1}\\
&\times[\textbf{B}_{i}\textbf{H}_d^H\textbf{W}\circ E(\textbf{y}_{i}\textbf{s}^H)^*\textbf{u}],
\end{split}
\end{equation}
where
\begin{equation}
\textbf{B}_{i} = \left\{ \begin{array}{ll}
\prod_{k=i+1}^{m-1}\textbf{H}_{k-1,k}^H\textbf{F}_k^H\textbf{A}_{k}^H, & \textrm{for $i=1,2,...,m-2$,}\\\textbf{I}, & \textrm{for $i=m-1$}.\\
\end{array}\right.
\end{equation}

The Lagrange multiplier $\lambda_i$ can be determined by solving
\begin{equation}
N_{i+1}\textbf{a}_{i,\rm opt}^H\textbf{a}_{i,\rm opt}=P_{T,i}~~i=1,2,...,m-1.
\end{equation}
Following the same steps as in Section III.A, we obtain $(m-1)$ $\{2M_i\}$th-order polynomials in $\lambda_i$
\begin{equation}
\sum_{j=1}^{M_i}N_{i+1}(\alpha_{i,j}+N_{i+1}\lambda_i)^{-2}\textbf{C}_i(j,j)=P_{T,i},~~i=1,2,...,m-1.
\end{equation}

\subsection{MMSE Design with Individual Power Constraints}
Thirdly, we consider the case where the power of each relay node is limited to some value $P_{T,i,j}$. The proposed method can be considered as the following optimization problem
\begin{equation}
\begin{split}
&[\textbf{W}_{\rm opt},\textbf{a}_{1,\rm opt},...,\textbf{a}_{m-1,\rm opt}] =\arg\min_{\textbf{W},\textbf{a}_1,...,\textbf{a}_{m-1}}E[\|\textbf{s}-\textbf{W}^H\textbf{d}\|^2],\\
&{\textrm {subject to}}  ~ P_{i,j}=P_{T,i,j},~i=1,2,...,m-1,~j=1,2,...,N_i,
\end{split}
\end{equation}
where $P_{i,j}$ is the transmitted power of the $j$th relay node in the $i$th group, and $P_{i,j}=N_{i+1}a_{i,j}^*a_{i,j}$.
Using the method of Lagrange multipliers once again, we have the following Lagrangian function
\begin{equation}
\mathcal{L}=E[\|\textbf{s}-\textbf{W}^H\textbf{d}\|^2] + \sum_{i=1}^{m-1}\sum_{j=1}^{N_i}\lambda_{i,j}(N_{i+1}a_{i,j}^*a_{i,j}-P_{T,i,j}). \\
\end{equation}
Following the same steps as described in Section III.A, we get the same optimal expression for $\textbf{W}$ as in (\ref{eq:mhop:MMSE:(2)}), and the optimal expression for the amplification coefficient
\begin{equation}
a_{i,j,\rm opt}=[\boldsymbol{\phi}_i(j,j)+N_{i+1}\lambda_{i,j}]^{-1}[\textbf{z}_i(j)-\sum_{l\in I, l\neq j}\boldsymbol{\phi}_i(j,l)a_{i,l}],
\end{equation}
where $I=\{1,2,...,N_i\}$, $\boldsymbol{\phi}_i$ and $\textbf{z}_i$ have the same expression as in (\ref{eq:mhop:MMSE:(7)}) and (\ref{eq:mhop:MMSE:(8)}). The Lagrange multiplier $\lambda_{i,j}$ can be determined by solving
\begin{equation}
N_{i+1}a_{i,j,\rm opt}^*a_{i,j,\rm opt}=P_{T,i,j}~~i=1,2,...,m-1,~~j=1,2,...N_i.
\end{equation}
Table I shows a summary of our proposed MMSE designs with global, local and individual power constraints which will be used for the simulations. If the quasi-static fading channel (block fading) is considered in the simulations, we only need two iterations. Alternatively, low-complexity adaptive algorithms can be used to compute the linear receiver $\textbf{W}_{\rm opt}$ and the power allocation parameter vector $\textbf{a}_{i,\rm opt}$.
\begin{table*}[!htb]
  \centering
  \caption{Summary of the Proposed MMSE Design with Global, local and individual Power Constraints}\label{}
  \begin{tabular}{p{5.5cm} | p{5.5cm} | p{5.5cm}}
  \hline
  ~~~Global Power Constraint& ~~~Local Power Constraints & ~~Individual Power Constraint\\
  \hline
  \hline
  Initialize the algorithm by setting: & Initialize the algorithm by setting: & Initialize the algorithm by setting:\\
  $ {\textbf{A}}=\sqrt{\frac{P_{T}}{\sum_{i=1}^{m-1}N_iN_{i+1}}}\textbf{I}$ & $ {\textbf{A}_i}=\sqrt{\frac{P_{T,i}}{N_iN_{i+1}}}\textbf{I} $ for $i=1,2,...,m-1$ & $a_{i,j}=\sqrt{\frac{P_{T,i,j}}{N_{i+1}}}$ for $i=1,2,...,m-1$, $j=1,2,...,N_i$\\
  For each iteration: & For each iteration: & For each iteration:\\
  1. Compute $\textbf{W}_{\rm opt}$ in (11). & 1. Compute $\textbf{W}_{\rm opt}$ in (11). & 1. Compute $\textbf{W}_{\rm opt}$ in (11). \\
  2. For $i=1,2,...,m-1$ & 2. For $i=1,2,...,m-1$  & 2. For $i=1,2,...,m-1$\\
  ~a) Compute $\boldsymbol{\phi}_i$ and $\textbf{z}_i$ in (21) and (22). & ~a) Compute $\boldsymbol{\phi}_i$ and $\textbf{z}_i$ in (21) and (22). & ~a) Compute $\boldsymbol{\phi}_i$ and $\textbf{z}_i$ in (21) and (22).\\
  ~b) Calculate the EVD of $\boldsymbol{\phi}_i$ in (26). & ~b) Calculate the EVD of $\boldsymbol{\phi}_i$ in (26).& ~b) For $j=1,2,...,N_i$\\
  ~c) Solve $\lambda$ in (31). & ~c) Solve $\lambda_i$ in (37). & ~~i) Solve $\lambda_{i,j}$ in (41).\\
  ~d) Compute $\textbf{a}_{i,\rm opt}$ in (18). & ~d) Compute $\textbf{a}_{i,\rm opt}$ in (34). & ~~ii) Compute $a_{i,j,\rm opt}$ in (40). \\
  \hline
\end{tabular}
\end{table*}

\section{Proposed Joint Maximum Sum-Rate Design of the Receiver and the Power Allocation}
In this section, we analyse the proposed joint MSR design of the receiver and the power allocation. By the MSR designs, the best possible SNR and QoS can be obtained at the destinations. They will improve the spectrum efficiency which is desirable for the WSNs with the limitation in the sensor node computational capacity. Only the local constraints are considered here, because of the MSR receiver we make use of the Generalized Rayleigh Quotient which is only suitable to solve optimization problems with vectors. It limits the types of power constraints.
By substituting (\ref{eq:mhop:MSR:(2)})-(\ref{eq:mhop:MSR:(7)}) into (\ref{eq:mhop:MSR:(8)}), we get
\begin{equation}
\begin{split}
\textbf{d} =& \textbf{C}_{0,m-1}\textbf{s} + \textbf{C}_{1,m-1}\textbf{v}_1 + \textbf{C}_{2,m-1}\textbf{v}_2\\
&+ ...+\textbf{C}_{m-1,m-1}\textbf{v}_{m-1}+\textbf{v}_d\\
=&\textbf{C}_{0,m-1}\textbf{s} + \sum_{i=1}^{m-1}\textbf{C}_{i,m-1}\textbf{v}_i+\textbf{v}_d,
\end{split}
\end{equation}
where
\begin{equation}
\textbf{C}_{i,j} = \left\{ \begin{array}{ll}
\prod_{k=i}^j\textbf{B}_k, & \textrm{if $i\leqslant j$,}\\\textbf{I}, & \textrm{if $i>j$},\\
\end{array}\right.
\end{equation}
and
\begin{equation}
\textbf{B}_0 = \textbf{H}_s,
\end{equation}
\begin{equation}
\textbf{B}_i = \textbf{H}_{i,i+1}\textbf{A}_i\textbf{F}_i~~~~~\textrm{for}~i=1,~2,~...,~m-2,
\end{equation}
\begin{equation}
\textbf{B}_{m-1} = \textbf{H}_d\textbf{A}_{m-1}\textbf{F}_{m-1}.
\end{equation}
We focus on a system with one source node for simplicity. The generalization to multiple sources amounts to performing the optimization of the additional filters. Therefore, the expression of the sum-rate (SR) in terms of bps/Hz for our $m$-hop WSN is expressed as
\begin{equation}
{\rm SR}=\frac{1}{m}\log_2\left[1+\frac{\sigma_s^2}{\sigma_n^2}\frac{\textbf{w}^H\textbf{C}_{0,m-1}\textbf{C}_{0,m-1}^H\textbf{w}}{\textbf{w}^H(\sum_{i=1}^m\textbf{C}_{i,m-1}\textbf{C}_{i,m-1}^H)\textbf{w}}\right]~\text{(bps/Hz)},
\end{equation}
where $\textbf{w}$ is the linear receiver, and $(\cdot)^H$ denotes the complex-conjugate (Hermitian) transpose. Let
\begin{equation}
\boldsymbol{\phi} = \textbf{C}_{0,m-1}\textbf{C}_{0,m-1}^H
\end{equation}
and
\begin{equation}
\textbf{Z} = \sum_{i=1}^m\textbf{C}_{i,m-1}\textbf{C}_{i,m-1}^H.
\end{equation}
The expression for the sum-rate can be written as
\begin{equation}
{\rm SR}=\frac{1}{m}\log_2\left(1+\frac{\sigma_s^2}{\sigma_n^2}\frac{\textbf{w}^H\boldsymbol{\phi}\textbf{w}}{\textbf{w}^H\textbf{Z}\textbf{w}}\right)=\frac{1}{m}\log_2(1+ax),
\end{equation}
where
\begin{equation}
a =\frac{\sigma_s^2}{\sigma_n^2}
\end{equation}
and
\begin{equation}
x =\frac{\textbf{w}^H\boldsymbol{\phi}\textbf{w}}{\textbf{w}^H\textbf{Z}\textbf{w}}.
\end{equation}
Since $\frac{1}{m}\log_2(1+ax)$ is a monotonically increasing function of $x$ ($a>0$), the problem of maximizing the sum-rate is equivalent to maximizing $x$. In this section, we consider the case where the total power of the relay nodes in each group is limited to some value $P_{T,i}$ (local constraints). The proposed method can be considered as the following optimization problem:
\begin{equation}
\label{eq:mhop:MSR:(20)}
\begin{split}
&[\textbf{w}_{\rm opt},\textbf{a}_{1,\rm opt},...,\textbf{a}_{m-1,\rm opt}] ~=\arg\max_{\textbf{w},\textbf{a}_1,...,\textbf{a}_{m-1}}\frac{\textbf{w}^H\boldsymbol{\phi}\textbf{w}}{\textbf{w}^H\textbf{Z}\textbf{w}},  \\
&{~~~~~~~~~\textrm {subject to}}  ~ P_i=P_{T,i},~i=1,2,...,m-1
\end{split}
\end{equation}
where $P_i$ as defined above is the transmitted power of the $i$th
group of relays, and $P_i=N_{i+1}\textbf{a}_i^H\textbf{a}_i$. We note that the expression
$\frac{\textbf{w}^H\boldsymbol{\phi}\textbf{w}}{\textbf{w}^H\textbf{Z}\textbf{w}}$
in (\ref{eq:mhop:MSR:(20)}) is the Generalized Rayleigh Quotient. Thus, the optimal
solution of our maximization problem can be obtained:
$\textbf{w}_{\rm opt}$ is any eigenvector corresponding to the dominant
eigenvalue of $\textbf{Z}^{-1}\boldsymbol{\phi}$.

In order to obtain the optimal power allocation vector
$\textbf{a}_{\rm opt}$,  we rewrite
$\frac{\textbf{w}^H\boldsymbol{\phi}\textbf{w}}{\textbf{w}^H\textbf{Z}\textbf{w}}$
and the expression is given by
\begin{equation}
\label{eq:mhop:MSR:(21)}
\frac{\textbf{w}^H{\boldsymbol \phi}\textbf{w}} {\textbf{w}^H
\textbf{Z} \textbf{w}} = \frac{\textbf{a}_i^H \textbf{M}_i
\textbf{a}_i} {\textbf{a}_i^H\textbf{P}_i\textbf{a}_i + \textbf{w}_i^H
\textbf{T}_i\textbf{w}_i}, ~{\rm for}~i=1,~2,...,~m-1,
\end{equation}
where
\begin{equation}
\begin{split}
\textbf{M}_i = &{\rm
diag}\{\textbf{w}_i^H\textbf{C}_{i+1,m-1}\textbf{H}_{i,i+1}\textbf{F}_i\}\textbf{C}_{0,i-1}\textbf{C}_{0,i-1}^H\\
&\times{\rm
diag}\{\textbf{F}_i^H\textbf{H}_{i,i+1}^H\textbf{C}_{i+1,m-1}^H\textbf{w}_i\},
\end{split}
\end{equation}
\begin{equation}
\begin{split}
\textbf{P}_i =&
{\rm
diag}\{\textbf{w}_i^H\textbf{C}_{i+1,m-1}\textbf{H}_{i,i+1}\textbf{F}_i\}(\sum_{k=1}^i\textbf{C}_{k,i-1}\textbf{C}_{k,i-1}^H)
\\
&\times{\rm
diag}\{\textbf{F}_i^H\textbf{H}_{i,i+1}^H\textbf{C}_{i+1,m-1}^H\textbf{w}_i\},
\end{split}
\end{equation}
and
\begin{equation}
\textbf{T}_i =
\sum_{k=i+1}^m\textbf{C}_{k,m-1}\textbf{C}_{k,m-1}^H.
\end{equation}
Since the multiplication of any constant value and an eigenvector is
still an eigenvector of the matrix, we express the receive
filter as
\begin{equation}
\textbf{w}_i =
\frac{\textbf{w}_{\rm opt}}{\sqrt{\textbf{w}_{\rm opt}^H\textbf{T}_i\textbf{w}_{\rm opt}}}.
\end{equation}
Hence, we obtain
\begin{equation}
\label{eq:mhop:MSR:(26)}
\textbf{w}_i^H\textbf{T}_i\textbf{w}_i = 1 = \frac{N_{i+1}\textbf{a}_i^H\textbf{a}_i}{P_{T,i}}.
\end{equation}
By substituting (\ref{eq:mhop:MSR:(26)}) into (\ref{eq:mhop:MSR:(21)}), we
get
\begin{equation}
\label{eq:mhop:MSR:(27)}
\frac{\textbf{w}^H\boldsymbol{\phi}\textbf{w}}{\textbf{w}^H\textbf{Z}\textbf{w}}
=
\frac{\textbf{a}_i^H\textbf{M}_i\textbf{a}_i}{\textbf{a}_i^H\textbf{N}_i\textbf{a}_i}~~~~{\rm
for}~i=1,~2,...,~m-1,
\end{equation}
where
\begin{equation}
\textbf{N}_i = \textbf{P}_i+\frac{N_{i+1}}{P_{T,i}}\textbf{I}.
\end{equation}
Likewise, we note that the expression
$\frac{\textbf{a}^H\textbf{M}_i\textbf{a}}{\textbf{a}_i^H\textbf{N}_i\textbf{a}_i}$
in (\ref{eq:mhop:MSR:(27)}) is the Generalized Rayleigh Quotient. Thus, the optimal
solution of our maximization problem can be obtained:
$\textbf{a}_{i,\rm opt}$ is any eigenvector corresponding to the
dominant eigenvalue of $\textbf{N}_i^{-1}\textbf{M}_i$ that
satisfies
$\textbf{a}_{i,\rm opt}^H\textbf{a}_{i,\rm opt}=\frac{P_{T,i}}{N_{i+1}}$.  Here, the local power constraints can be satisfied by employing a normalization. When considering the global power constraint $P_T$, there is no unique solution of $\textbf{a}_{i,\rm opt}$ $(i=1,2,...,m-1)$ that satisfy $\sum_{i=1}^{m-1}N_{i+1}\textbf{a}_{i,\rm opt}^H\textbf{a}_{i,\rm opt}=P_T$. Thus, for this reason, we do not consider the global power constraint.
The solutions of $\textbf{w}_{\rm opt}$ and $\textbf{a}_{i,\rm opt}$ depend
on each other. Therefore it is necessary to iterate them with an
initial value of $\textbf{a}_i$ ($i=1,2,...,m-1$) to obtain the
optimum solutions.

In this section, two methods are employed to calculate the dominant
eigenvectors. The first one is the QR algorithm \cite{Watkins} which
calculates all the eigenvalues and eigenvectors of a matrix. We can
choose the dominant eigenvector among them. The second one is the
power method \cite{Watkins} which only calculates the dominant
eigenvector of a matrix. Hence, the computational complexity can
be reduced. Table II shows a summary of our proposed MSR design with a local power constraint which will be used for the simulations. If the quasi-static fading channel (block fading) is considered in the simulations, we only need two iterations.
\begin{table}[!htb]
  \centering
  \caption{Summary of the Proposed MSR Design with Local Power Constraints}\label{}
  \begin{tabular}{l}
  \hline
Initialize the algorithm by setting \\
~~~~~~~~~~~~$ {\textbf{A}_i}=\sqrt{\frac{P_{T,i}}{N_iN_{i+1}}}\textbf{I} $ ~~for $i=1,2,...,m-1$\\
For each iteration:\\
1. Compute $\boldsymbol{\phi}$ and $\textbf{Z}$ in (48) and (49).\\
2. Using the QR algorithm or the power method to\\~~~compute the dominant eigenvector of $\textbf{Z}^{-1}\boldsymbol{\phi}$, \\~~~denoted as $\textbf{w}_{\rm opt}$.\\

3. For $i = 1,2,...,m-1$\\
~~~a) Compute $\textbf{M}_i$ and $\textbf{N}_i$ in (55) and (61).\\
~~~b) Using the QR algorithm or the power method to\\~~~~~~compute the dominant eigenvector of $\textbf{N}_i^{-1}\textbf{M}_i$,\\~~~~~~denoted as $\textbf{a}_{i}$.\\
~~~c) To ensure the local power constraint\\~~~~~~ $\textbf{a}_{i,\rm opt}^H\textbf{a}_{i,o\rm pt}=\frac{P_{T,i}}{N_{i+1}}$, compute $\textbf{a}_{i,\rm opt}=\sqrt{\frac{P_{T,i}}{N_{i+1}\textbf{a}_i^H\textbf{a}_i}}\textbf{a}_{i}$.  \\
\hline
\end{tabular}
\end{table}

\section{Analysis of the proposed algorithms}
In this section, an analysis of the computational complexity and the convergence of the algorithms is developed. We first illustrate the computational complexity requirements of the proposed MMSE and MSR designs. We quantify the computational complexity of the algorithms, which require a given number of arithmetic operations per iteration. The lower the number of operations the lower the power consumption will be. Then, we make use of the convergence results for the alternating optimization algorithms in \cite{Csiszar,Niesen} and present a set of sufficient conditions under which our proposed algorithms will converge to the optimal solutions.
\subsection{Computational Complexity Analysis}
Table III and Table IV list the computational complexity per iteration in terms of the number of multiplications, additions and divisions for our proposed joint linear receiver design (MMSE and MSR) and power allocation strategies. For the joint MMSE designs, we use the QR algorithm to perform the eigendecomposition of the matrix. Please note that in this paper the QR decomposition employs the Householder transformation \cite{Watkins,Golub}. The quantities $n_Q$ and $n_P$ denote the number of iterations of the QR algorithm and the power method, respectively. For the computational complexity of $\lambda$ in Table III, it does not include the processing of solving the equation in (31), (37) and (41), because the method with a global power constraint, equation (31) is a higher order polynomial whose complexity is difficult to be quantified. As the multiplication dominates the computational complexity, in order to compare the computational complexity of our proposed joint MMSE and MSR designs,  the number of multiplications versus the number of relay nodes in each group for each iteration are displayed in Fig. 3 and Fig 4. For the purpose of illustration, we set $m=3$, $N_0=1$, $N_3=2$ and $n_Q = n_P = 10$. For the MMSE design, it can be seen that our proposed receiver with a global constraint has the same complexity as the receiver with local constraints. In practice, when considering the processing of solving the equation in (31), (37), the method with a global constraint will require a higher computational complexity than the local constraints and the difference will become larger along with the increase of the number of hops ($m$).  When the individual power constraints are considered, the computational complexity is lower than other constraints because there is no need to compute the eigendecomposition for it. For the MSR design, employing the power method to calculate the dominant eigenvectors has a lower computational complexity than employing the QR algorithm.

\section{Analysis of the proposed algorithms}
In this section, an analysis of the computational complexity and the convergence of the algorithms is developed. We first illustrate the computational complexity requirements of the proposed MMSE and MSR designs. Then, we make use of the convergence results for the alternating optimization algorithms in \cite{Csiszar,Niesen} and present a set of sufficient conditions under which our proposed algorithms will converge to the optimal solutions.
\subsection{Computational Complexity Analysis}
Table III and Table IV list the computational complexity per iteration in terms of the number of multiplications, additions and divisions for our proposed joint linear receiver design (MMSE and MSR) and power allocation strategies. For the joint MMSE designs, we use the QR algorithm to perform the eigendecomposition of the matrix. Please note that in this paper the QR decomposition by the Householder transformation \cite{Watkins,Golub} is employed by the QR algorithms. The quantities $n_Q$ and $n_P$ denote the number of iterations of the QR algorithm and the power method, respectively. For the computational complexity of $\lambda$ in Table III, it does not include the processing of solving the equation in (31), (37) and (41), because of the method with a global power constraint, equation (31) is a higher order polynomial whose complexity is difficult to be summarized. As the multiplication dominates the computational complexity, in order to compare the computational complexity of our proposed joint MMSE and MSR designs,  the number of multiplications versus the number of relay nodes in each group for each iteration are displayed in Fig. 3 and Fig 4. For the purpose of illustration, we set $m=3$, $N_0=1$, $N_3=2$ and $n_Q = n_P = 10$. For the MMSE design, it can be seen that our proposed receiver with a global constraint has the same complexity as the receiver with local constraints. In practice, when considering the processing of solving the equation in (31), (37), the method with a global constraint will require a higher computational complexity than the local constraints and the difference will become larger along with the increase of the number of hops ($m$).  When the individual power constraints are considered, the computational complexity is lower than other constraints because there is no need to compute the eigendecomposition for it. For the MSR design, employing the power method to calculate the dominant eigenvectors has a lower computational complexity than employing the QR algorithm.
\begin{table*}[!htb]
  \centering
\begin{footnotesize}
  \centering
  \caption{Computational Complexity per Iteration of the joint MMSE Designs }\label{}
  \begin{tabular}{c c c c c}
  \hline
  &Power Constraint& Multiplications & Additions & Divisions\\
  \hline
  \\
               && $N_m(N_m-1)(4N_m+1)/6$ & $N_m(N_m-1)(4N_m+1)/6$ &  \\
  && $+(N_0+N_{m-1})N_d^2+N_{m-1}^2N_m$ & $+(N_0+N_{m-1})N_m^2+N_{m-1}^2N_m$ &  \\
    $\textbf{W}$            &All& $+N_0N_{m-1}N_m+N_{m-1}N_m$ & $+N_0N_{m-1}N_m-N_m^2+2N_0N_m$ & $N_m(3N_m-1)/2$\\
               && $+\sum_{i=2}^{m-1}\{2N_{i-1}^2N_i+N_{i-1}N_i^2$ & $+N_{m-1}N_m+N_m+\sum_{i=2}^{m-1}\{2N_{i-1}N_i^2$ & \\
               && $+N_0N_{i-1}N_i+4N_{i-1}N_i+2N_i\}$ & $+N_0(N_{i-1}-1)N_i-N_i^2+N_i\}$ & \\
  \\
  \hline
  \\
            && $\sum_{i=1}^{m-1}\{n_Q(\frac{13}{6}N_i^3+\frac{3}{2}N_i^2+\frac{1}{3}N_i-2)$ & $\sum_{i=1}^{m-1}\{n_Q(\frac{13}{6}N_i^3-N_i^2-\frac{1}{6}N_i+1)$ & \\
            &Global& $-N_i^3+3N_0N_i^2+N_0N_iN_{i+1}+N_i^2\}$ & $-N_i^3+3N_0N_i^2+N_0N_iN_{i+1}$ & $\sum_{i=1}^{m-1}\{n_Q(N_i-1)\}$\\
            && $+\sum_{i=1}^{m-2}\{N_iN_{i+1}+N_{i+1}\}$ & $-N_i^2-N_0N_i-N_i\}$ & \\
  \\
            && $\sum_{i=1}^{m-1}\{n_Q(\frac{13}{6}N_i^3+\frac{3}{2}N_i^2+\frac{1}{3}N_i-2)$ & $\sum_{i=1}^{m-1}\{n_Q(\frac{13}{6}N_i^3-N_i^2-\frac{1}{6}N_i+1)$ & \\
 $\lambda$ &Local& $-N_i^3+3N_0N_i^2+N_0N_iN_{i+1}+N_i^2\}$ & $-N_i^3+3N_0N_i^2+N_0N_iN_{i+1}$ & $\sum_{i=1}^{m-1}\{n_Q(N_i-1)\}$\\
            && $+\sum_{i=1}^{m-2}\{N_iN_{i+1}+N_{i+1}\}$ & $-N_i^2-N_0N_i-N_i\}$ & \\
  \\
    &Individual& $\sum_{i=1}^{m-1}\{N_0N_i^2+N_0N_iN_{i+1}+N_i^2+N_0N_i\}$ & $\sum_{i=1}^{m-1}\{N_0N_i^2+N_0N_iN_{i+1}-N_i^2-N_i\}$ & \\
      &&$+\sum_{i=1}^{m-2}\{N_iN_{i+1}+N_{i+1}\}$ &  & \\
   \\
  \hline
  \\
  &Global& $\sum_{i=1}^{m-1}\{N_i(N_i-1)(4N_i+1)/6+N_i^2+1\}$ & $\sum_{i=1}^{m-1}\{N_i(N_i-1)(4N_i+1)/6+N_i^2\}$ & $\sum_{i=1}^{m-1}\{N_i(3N_i-1)/2\}$\\
  \\
  $\textbf{a}$&Local& $\sum_{i=1}^{m-1}\{N_i(N_i-1)(4N_i+1)/6+N_i^2+1\}$ & $\sum_{i=1}^{m-1}\{N_i(N_i-1)(4N_i+1)/6+N_i^2\}$ & $\sum_{i=1}^{m-1}\{N_i(3N_i-1)/2\}$\\
  \\
              &Individual& $2\sum_{i=1}^{m-1}N_i$ & $\sum_{i=1}^{m-1}N_i$ & $\sum_{i=1}^{m-1}N_i$\\
  \\
  \hline
\end{tabular}
\end{footnotesize}
\end{table*}

\begin{table*}[!htb]
  \centering
\begin{footnotesize}
  \centering
  \caption{Computational Complexity per Iteration of the joint MSR Designs}\label{}
  \begin{tabular}{c c c c c}
  \hline
  &Power Constraint& Multiplications & Additions & Divisions\\
  \hline
  \\
           &  & $n_Q(\frac{13}{6}N_m^3+\frac{3}{2}N_m^2+\frac{1}{3}N_m-2)$ & $n_Q(\frac{13}{6}N_d^3-N_d^2-\frac{1}{6}N_d+1)$ & \\
  &  Local& $+N_m(N_m-1)(4N_m+1)/6+N_m^2+N_1N_m$ & $+N_m(N_m-1)(4N_m+1)/6$ & $n_Q(N_m-1)$ \\
     &QR Algorithm& $+\sum_{i=1}^{m-1}\{N_iN_m^2+N_iN_{i+1}+N_i\}$ & $-N_m^2+N_1N_m+\sum_{i=1}^{m-1}N_iN_m^2$ & $+N_m(3N_m-1)/2$ \\
                && $+\sum_{i=2}^{m-1}\{2N_{i-1}^2N_i+N_{i-1}N_i^2$ & $+\sum_{i=2}^{m-1}\{2N_{i-1}N_i^2$ & \\
                && $+N_{i-1}N_iN_m+4N_{i-1}N_i+2N_i\}$ & $+N_{i-1}(N_i-1)N_m-N_i^2+N_i\}$ & \\
  $\textbf{w}$ &&&\\
               && $n_PN_m^2+N_m(N_m-1)(4N_m+1)/6$ & $n_PN_m(N_m-1)$ &\\
  &Local& $+N_m^3+N_m^2+N_1N_m$ & $+N_m(N_m-1)(4N_m+1)/6+N_m^3$ &  \\
             &Power Method& $+\sum_{i=1}^{m-1}\{N_iN_m^2+N_iN_{i+1}+N_i\}$ & $-2N_m^2+N_1N_m+\sum_{i=1}^{m-1}N_iN_m^2$ & $N_m(3N_m-1)/2$ \\
                && $+\sum_{i=2}^{m-1}\{2N_{i-1}^2N_i+N_{i-1}N_i^2$ & $+\sum_{i=2}^{m-1}\{2N_{i-1}N_i^2$ & \\
                && $+N_{i-1}N_iN_m+4N_{i-1}N_i+2N_i\}$ & $+N_{i-1}(N_i-1)N_m-N_i^2+N_i\}$ & \\
  \\
  \hline
  \\
               && $\sum_{i=1}^{m-1}\{n_Q(\frac{13}{6}N_i^3+\frac{3}{2}N_i^2+\frac{1}{3}N_i-2)$ & $\sum_{i=1}^{m-1}\{n_Q(\frac{13}{6}N_i^3-N_i^2-\frac{1}{6}N_i+1)$ & \\
  &Local& $+N_i(N_i-1)(4N_i+1)/6$ & $+N_i(N_i-1)(4N_i+1)/6+N_iN_{i+1}$ & $\sum_{i=1}^{m-1}\{n_Q(N_i-1)$\\
               &QR Algorithm& $+\sum_{k=1}^iN_kN_i^2+3N_i^2+2N_iN_{i+1}$ & $+N_{i+1}N_m-N_{i+1}+N_i-1\}$ & $+N_i(3N_i-1)/2\}$  \\
                           && $+N_{i+1}N_m+3N_i+2\}+2N_m^2$ & $+\sum_{i=2}^{m-1}\{\sum_{k=1}^{i-1}(N_k-1)N_i^2$ & $+N_m+m-1$ \\
                          $\textbf{a}$ &&  & $+N_i^2(i-2)+N_i\}+2N_m^2-2N_m$ &  \\
  \\
               && $\sum_{i=1}^{m-1}\{n_PN_i^2$ & $\sum_{i=1}^{m-1}\{n_PN_r(N_r-1)$ & \\
  &Local& $+N_i(N_i-1)(4N_i+1)/6$ & $+N_i(N_i-1)(4N_i+1)/6+N_i^3$ & $\sum_{i=1}^{m-1}\{N_i(3N_i-1)/2\}$ \\
                           &Power Method& $+\sum_{k=1}^iN_kN_i^2+N_i^3+3N_i^2+2N_iN_{i+1}$ & $-N_i^2+N_iN_{i+1}+N_{i+1}N_m-N_{i+1}$ & $+N_m+m-1$ \\
                           && $+N_{i+1}N_m+3N_i+2\}+2N_m^2$ & $+N_i-1\}+\sum_{i=2}^{m-1}\{\sum_{k=1}^{i-1}(N_k-1)N_i^2$ &  \\
                           &&  & $+N_i^2(i-2)+N_i\}+2N_m^2-2N_m$ &  \\
  \\
  \hline
\end{tabular}
\end{footnotesize}
\end{table*}

\begin{figure}[!htb]
\centering
\includegraphics[width=3.5in]{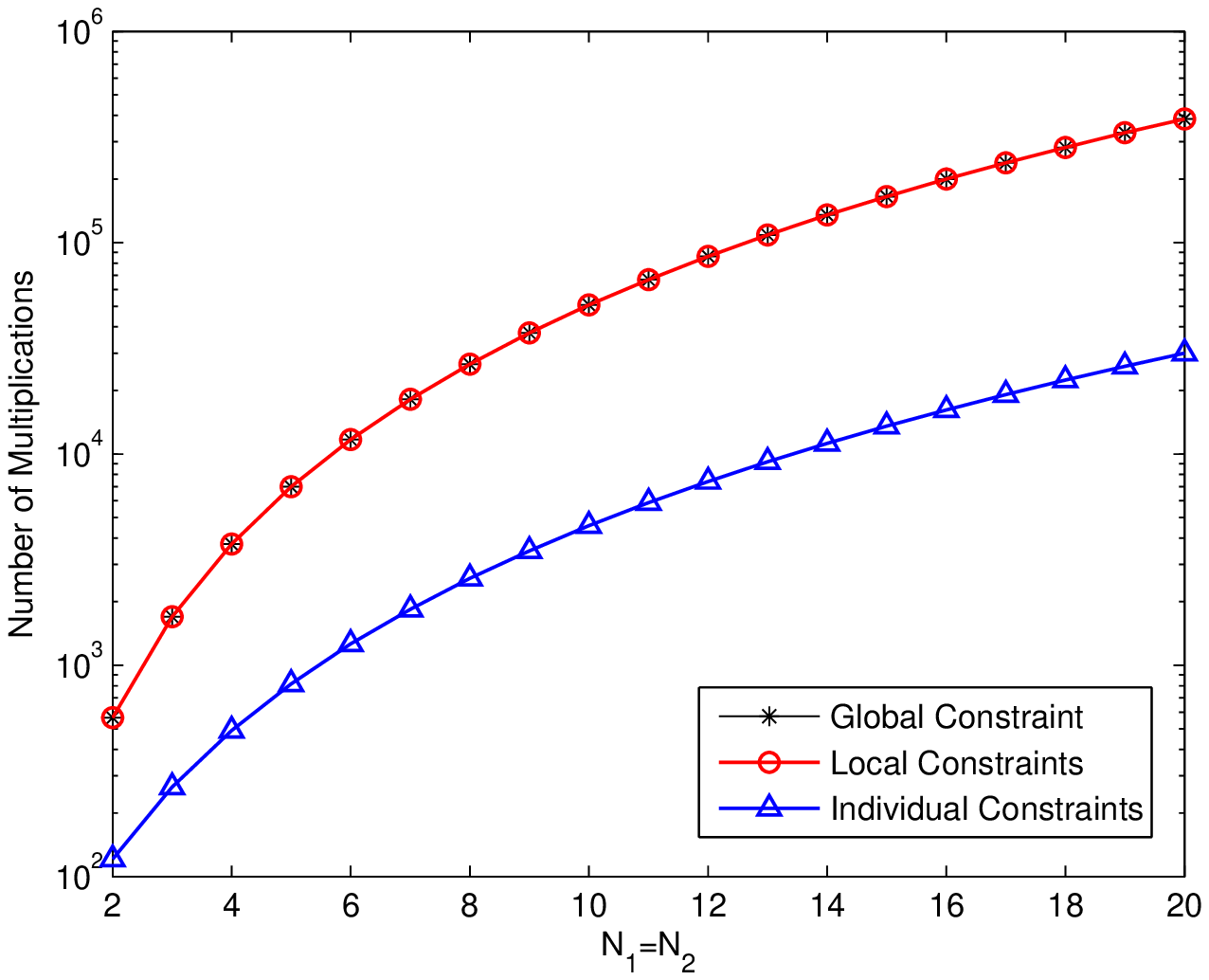}
\caption{Number of multiplications versus the number of relay nodes of our proposed joint MMSE design of the receiver and the power allocation strategies.}
\end{figure}

\begin{figure}[!htb]
\centering
\includegraphics[width=3.5in]{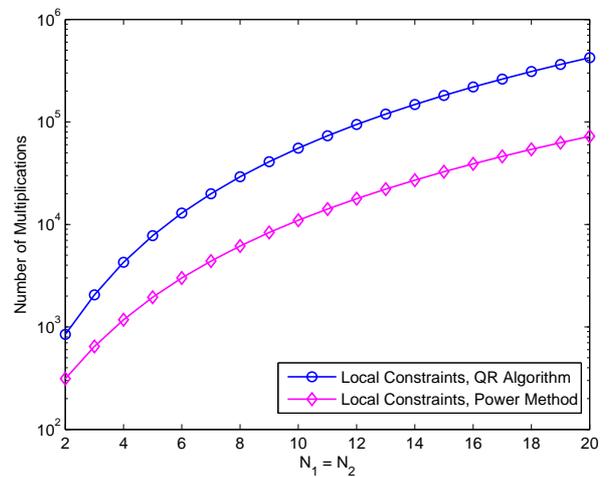}
\caption{Number of multiplications versus the number of relay nodes of our proposed joint MSR design of the receiver and the power allocation strategies.}
\end{figure}

\subsection{Sufficient Conditions for Convergence}
To obtain convergence conditions, we need to define a metric space and the Hausdorff distance that will extensively be used. A metric space is an ordered pair $(\mathcal{M}, d)$, where $\mathcal{M}$ is a nonempty set, and $d$ is a metric on $\mathcal{M}$, i.e., a function $d : \mathcal{M}\times \mathcal{M}\rightarrow \mathbb{R}$ such that for any $x, y, z$ $\in$ $\mathcal{M}$, the following
conditions hold:

1) $d(x,y)\geq0$.

2) $d(x,y)=0~ \text{iff}~ x=y$.

3) $d(x,y)=d(y,x)$.

4) $d(x,y)\leq d(x,y)+d(y,z)$.

The Hausdorff distance measures how far two subsets of a metric space are from each other and is defined by
\begin{equation}
d_H(X,Y)={\rm max}\left\{\sup_{x\in X} \inf_{y\in Y}d(x,y),\sup_{y\in Y} \inf_{x\in X}d(x,y)\right\}.
\end{equation}

The proposed joint MMSE designs can be stated as an alternating minimization strategy based on the MSE defined in (9) and expressed as
\begin{equation}
\textbf{W}_n\in \arg \min_{\textbf{W}\in \underline{\textbf{W}}_n}{\rm MSE}(\textbf{W},\textbf{a}_{i,n-1}),
\end{equation}
\begin{equation}
\textbf{a}_{i,n}\in \arg \min_{\textbf{a}\in \underline{\textbf{a}}_{i,n}}{\rm MSE}(\textbf{W}_{n},\textbf{a}_i)~~\text{for}~ i=1,2,...,m-1
\end{equation}
where the sets $\underline{\textbf{W}},\underline{\textbf{a}}_{i}\subset\mathcal{M}$, and the sequences of compact
sets $\{\underline{\textbf{W}}_n\}_{n\geq 0}$ and $\{\underline{\textbf{a}}_{i,n}\}_{n\geq 0}$ converge to the sets $\underline{\textbf{W}}$ and $\underline{\textbf{a}}_{i}$, respectively.

Although we are not given the sets $\underline{\textbf{W}}$ and $\underline{\textbf{a}}_i$ directly, we have the sequence of compact sets $\{\underline{\textbf{W}}_n\}_{n\geq 0}$ and $\{\underline{\textbf{a}}_{i,n}\}_{n\geq 0}$. The aim of our proposed joint MMSE designs  is to find a sequence of $\textbf{W}_{n}$ and $\textbf{a}_{i,n}(i=1,2,...,m-1)$ such that
\begin{equation}
\begin{split}
&\lim_{n\rightarrow \infty}{\rm MSE}(\textbf{W}_{n},\textbf{a}_{1,n},\textbf{a}_{2,n},...,\textbf{a}_{m-1,n})\\
&={\rm MSE}(\textbf{W}_{opt},\textbf{a}_{1,opt},\textbf{a}_{2,opt},...,\textbf{a}_{m-1,opt})
\end{split}
\end{equation}
where $\textbf{W}_{opt}$ and $\textbf{a}_{i,opt}$ correspond to the optimal values of $\textbf{W}_n$ and $\textbf{a}_{i,n}$, respectively. Equation (65) can be considered as
the necessary condition of the following equations
\begin{equation}
\begin{split}
&\lim_{n\rightarrow \infty}{\rm MSE}(\textbf{W}_{n},\textbf{a}_{i,n})={\rm MSE}(\textbf{W}_{opt},\textbf{a}_{i,opt}) \\
&~~~~~~~~\text{for}~ i=1,2,...,m-1
\end{split}
\end{equation}
if the other power allocation parameters $\textbf{a}_{j,n} (j\neq i)$ are kept constant when computing $\textbf{a}_{i,n}$ during the iterations.
To present a set of sufficient conditions under which the proposed algorithms converge, we need the so-called three-point and four-point properties \cite{Csiszar,Niesen}. Let us assume that there is a function $f : \mathcal{M}\times \mathcal{M}\rightarrow \mathbb{R}$ such that the following conditions are satisfied:
\begin{description}

\item[1)] \emph{Three-point property} $(\textbf{W},\widetilde{\textbf{W}},\textbf{a}_i)$:\\For all $n\geq 1$, $\textbf{W} \in \underline{\textbf{W}}_n$, $\textbf{a}_i \in \underline{\textbf{a}}_{i,n-1}$, and \\$\widetilde{\textbf{W}}\in \arg\min_{\textbf{W}\in \underline{\textbf{W}}_n}{\rm MSE}(\textbf{W},\textbf{a}_i)$, we have
\begin{equation}
f(\textbf{W},\widetilde{\textbf{W}}) + {\rm MSE}(\widetilde{\textbf{W}},\textbf{a}_i)\leq {\rm MSE}(\textbf{W},\textbf{a}_i).
\end{equation}

\item[2)] \emph{Four-point property} $(\textbf{W},\textbf{a}_i,\widetilde{\textbf{W}},\tilde{\textbf{a}}_i)$:\\ For all $n\geq 1$, $\textbf{W},\widetilde{\textbf{W}} \in \underline{\textbf{W}}_n$, $\textbf{a}_i \in \underline{\textbf{a}}_{i,n}$, and \\$\tilde{\textbf{a}}_i\in \arg \min_{\textbf{a}_i\in \underline{\textbf{a}}_{i,n}}{\rm MSE}(\widetilde{\textbf{W}},\textbf{a}_i)$, we have
\begin{equation}
 {\rm MSE}(\textbf{W},\tilde{\textbf{a}}_i)\leq{\rm MSE}(\textbf{W},\textbf{a}_i) + f(\textbf{W},\widetilde{\textbf{W}}).
\end{equation}
\end{description}
These two properties are the mathematical expressions of the sufficient conditions for the convergence of the alternating minimization algorithms which are stated in \cite{Csiszar} and \cite{Niesen}. It means that if there exists a function $f(\textbf{W},\widetilde{\textbf{W}})$ with the parameter $\textbf{W}$ during two iterations that satisfies the two inequalities for the MSE in (67) and (68), the convergence of our proposed MMSE designs that make use of the alternating minimization algorithm can be proved by the theorem below.

\emph{Theorem}: Let $\{(\underline{\textbf{W}}_n,\underline{\textbf{a}}_{i,n})\}_{n\geq 0}$, $\underline{\textbf{W}},\underline{\textbf{a}}_i$ be compact subsects of the compact metric space $(\mathcal{M}, d)$ such that
\begin{equation}
\underline{\textbf{W}}_n \overset{d_H}\rightarrow \underline{\textbf{W}}~~~~~\underline{\textbf{a}}_{i,n} \overset{d_H}\rightarrow \underline{\textbf{a}}_i
\end{equation}
and let MSE : $\mathcal{M}\times \mathcal{M}\rightarrow \mathbb{R}$ be a continuous function. Let conditions 1) and 2) hold. Then, for the proposed algorithms, we have
\begin{equation}
\begin{split}
&\lim_{n\rightarrow \infty}{\rm MSE}(\textbf{W}_{n},\textbf{a}_{i,n})={\rm MSE}(\textbf{W}_{opt},\textbf{a}_{i,opt})\\
&~~~~~~~~\text{for}~ i=1,2,...,m-1.
\end{split}
\end{equation}
Thus, equation (65) can be satisfied.
A general proof of this theorem is detailed in \cite{Csiszar} and \cite{Niesen}. The proposed joint MSR designs can be stated as an alternating maximization strategy based on the SR defined in (47) that follows a similar procedure to the one above.
\section{simulations}
In this section, we assess the performance of our proposed joint designs of the linear receiver and power allocation methods and compare them with the equal power allocation method which allocates the same transmitting power level equally for all links from the relay nodes. For the purpose of fairness, we assume that the total transmitting power for all relay nodes in the network is the same which can be indicated as $\sum_{i=1}^{m-1}P_{T,i}=\sum_{i=1}^{m-1}\sum_{j=1}^{N_i}P_{T,i,j}=P_T$. We consider a 3-hop ($m$=3) wireless sensor
network as an example even though the algorithms can be used with any number of hops. The number of source nodes ($N_0$), two groups of relay nodes
($N_1, N_2$) and destination nodes ($N_3$) are 1, 4, 4 and 2,
respectively. We consider an AF cooperation protocol. The quasi-static fading channel (block fading channel)
is considered in our simulations whose elements are Rayleigh
random variables (with zero mean and unit variance) and assumed to be invariant during the
transmission of each packet. In our simulations, the channel is assumed to be known at the destination nodes. For channel estimation algorithms for WSNs and other low-complexity parameter estimation algorithms, one refers to \cite{Wang} and \cite{Lamare1}. During each phase, the sources
transmit the QPSK modulated packets with
1500 symbols. The noise at the
destination nodes is modeled as circularly symmetric complex
Gaussian random variables with zero mean. A perfect (error free) feedback channel between destination nodes and relay nodes is assumed to transmit the amplification coefficients.

For the MMSE design, it can be seen from Fig. 5 that our three proposed methods achieve a better performance than the equal power allocation method. Among them, the method with a global constraint has the best performance whereas the method with individual constraints has the worst performance. This result is what we expect because a global constraint provides the largest degrees of freedom for allocating the power among the relay nodes whereas an individual constraint provides the least. For the MSR design, it can be seen from Fig. 6 that our proposed method achieves a better sum-rate performance than the equal power allocation method. Using the power method to calculate the dominant eigenvector yields a very similar result to the QR algorithm but requires a lower complexity.
\begin{figure}[!htb]
\centering
\includegraphics[width=3.5in]{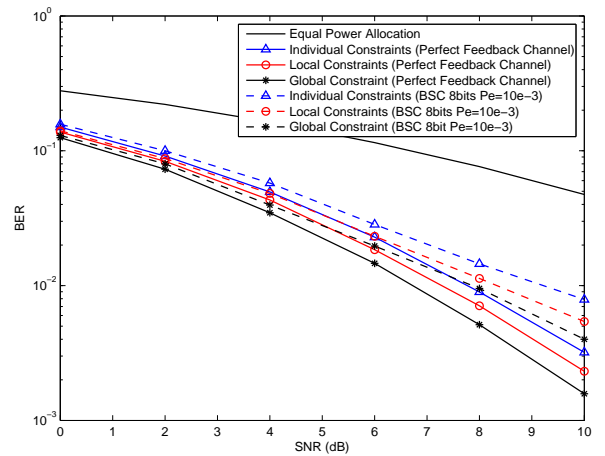}
\caption{BER performance versus SNR of our proposed joint MMSE design of the receiver and power allocation strategies, compared to the equal power allocation method.}
\end{figure}
\begin{figure}[!htb]
\centering
\includegraphics[width=3.5in]{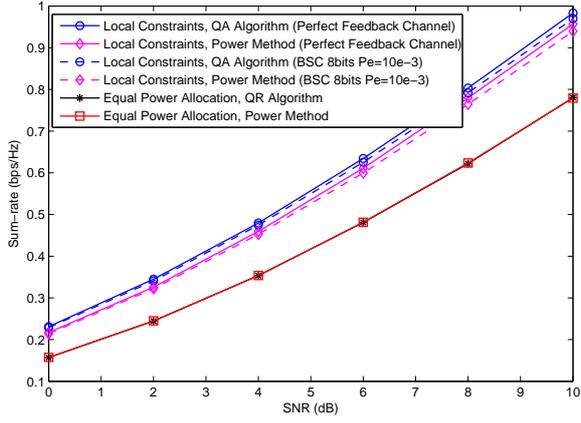}
\caption{Sum-rate performance versus SNR of our proposed joint MSR design of the receiver and power allocation strategies with local constraints, compared to the equal power allocation method.}
\end{figure}

Besides the equal power allocation scheme, a MMSE power allocation scheme reported in \cite{Rong} where only the local power constraints are considered has also been used for comparison. It can be seen from Fig. 7 that our proposed MMSE and MSR designs can achieve a very similar or better performance. Further advantage is that our proposed schemes only optimize the relay amplifying vectors (or diagonal matrices) whereas in \cite{Rong} the optimal relay amplifying matrices are needed which requires more feedback transmissions as well as information exchanges among relay nodes in each group. Note that in order to have a fair comparison, we only employ power allocation schemes for the relay nodes and assume every source node has unit transmitting power in the simulations.
\begin{figure}[!htb]
\centering
\includegraphics[width=3.5in]{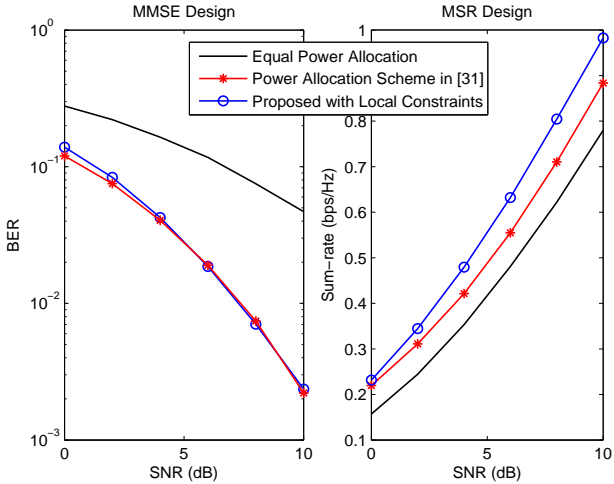}
\caption{(a) BER performance versus SNR of our proposed MMSE design (b) Sum-rate performance versus SNR of our proposed MSR design with local power constraints and compare with the power allocation scheme in [31] and equal power allocation scheme.}
\end{figure}

In practice, the feedback channel cannot be error free. In order to study the impact of feedback channel errors on the performance, we employ the binary symmetric channel (BSC) as the model for the feedback channel and quantize each complex amplification coefficient to an 8-bit binary value (4 bits for the real part, 4 bits for the imaginary part). The error probability (Pe) of the BSC is fixed at $10^{-3}$. The dashed curves in Fig. 5 and Fig. 6 show the performance degradation compared to the performance when using a perfect feedback channel. To show the performance tendency of the BSC for other values of Pe, we fix the SNR at 10 dB and choose Pe ranging from 0 to $10^{-2}$. The performance curves are shown in Fig. 8 and Fig. 9 , which illustrate the BER and the sum-rate performance versus Pe of our two proposed joint designs of the receivers. It can be seen that along with the increase in Pe, their performance becomes~ worse.
\begin{figure}[!htb]
\centering
\includegraphics[width=3.5in]{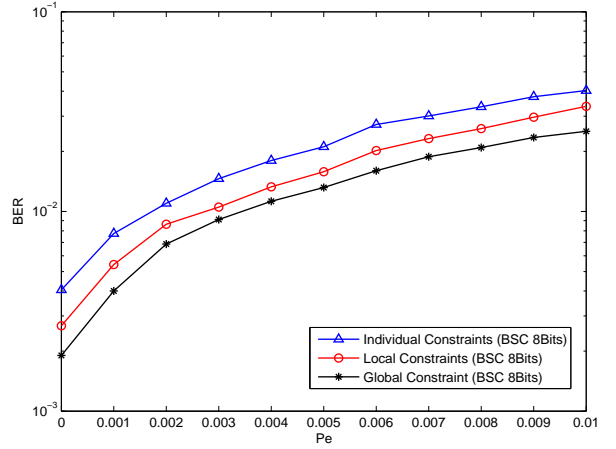}
\caption{BER performance versus Pe of our proposed MMSE designs when employing the BSC as the model for the feedback channel.}
\end{figure}

\begin{figure}[!htb]
\centering
\includegraphics[width=3.5in]{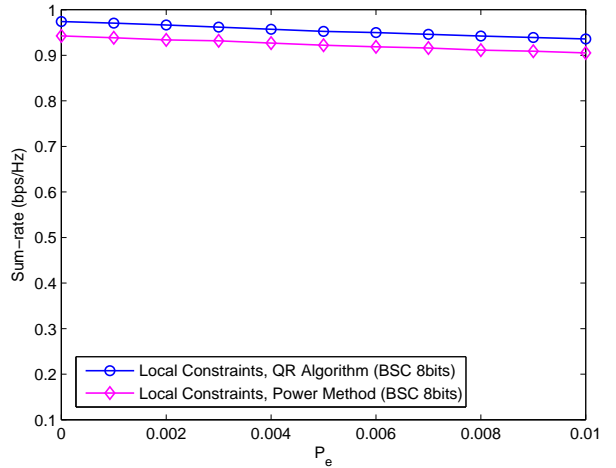}
\caption{Sum-rate performance versus Pe
of our proposed MSR design when employing the BSC as the model for
the feedback channel.}
\end{figure}

Finally, we replace the perfect CSI with the estimated channel coefficients to compute the receive filters and power allocation parameters at the destinations. We employ the BEACON channel estimation which was proposed in \cite{Wang}. Fig. 10 illustrates the impact of the channel estimation on the performance of our proposed MMSE and SMR design with local constraints by comparing it to the performance of perfect CSI. The quantity $n_t$ denotes the number of training sequence symbols per data packet. Please note that in these simulations perfect feedback channel is considered and the QR algorithm is used in the MSR design. For both the MMSE and MSR designs, it can be seen that when $n_t$ is set to 10, the BEACON channel estimation leads to an obvious performance degradation compared to the perfect CSI. However, when $n_t$ is increased to 50, the BEACON channel estimation can achieve a similar performance to the perfect CSI. Other scenarios and network topologies have been investigated and the results show that the proposed algorithms work very well with channel estimation algorithms and a small number of training~symbols.
\begin{figure}[!t]
\centering
\includegraphics[width=3.5in]{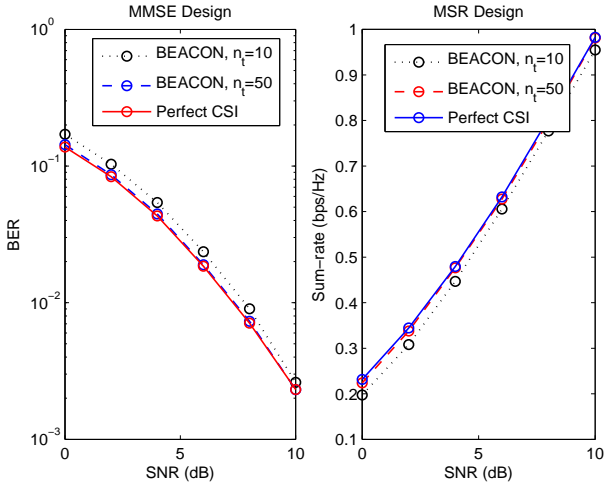}
\caption{(a) BER performance versus SNR of our proposed MMSE design (b) Sum-rate performance versus SNR our proposed MSR design with local power constraints when employing the BEACON channel estimation, compared to the performance of perfect CSI}
\end{figure}

\section{Conclusions}
In this paper, we have presented alternating optimization algorithms for receive filter design and power adjustment which can be applied to general multihop WSNs. MMSE and MSR criteria have been considered in the development of the algorithmic solutions. Simulations have shown that our proposed algorithms achieve a significant better performance than the equal power allocation and power allocation scheme in \cite{Rong}. A possible extension of this work is employing low-complexity adaptive algorithms to compute the linear receiver and power allocation parameters. The algorithms can also be employed in other multihop wireless networks along with non-linear receivers.


\appendix

Here, we derive the the expressions of $\textbf{F}_i$, $E(\textbf{y}_{i}\textbf{y}_{i}^H)$, and $E(\textbf{y}_{i}\textbf{s}^H)$ that are used in Section II, III and VI. It holds that
\begin{equation}
\textbf{F}_i={\rm diag}\{E(|x_{i,1}|^2),E(|x_{i,2}|^2),...,E(|x_{i,N_i}|^2)\}^{-\frac{1}{2}}
\end{equation}
where
\begin{equation}
E(|x_{i,j}|^2 = \left\{ \begin{array}{ll}
\sigma_s^2|\textbf{h}_{s,j}|^2+\sigma_n^2, ~~~~~~~~~~~~~~~~~~~~~~~\textrm{for $i=1$,}\\\textbf{h}_{i-1,j}\textbf{A}_{i-1}E(\textbf{y}_{i-1}\textbf{y}_{i-1}^H)\textbf{A}_{i-1}^H\textbf{h}_{i-1,j}^H+\sigma_n^2,\\~~~~~~~~~~~~~~~~~~~~~~~~~~~~~~~~\textrm{for $i=2,3,...,m$},\\
\end{array}\right.
\end{equation}

\begin{equation}
E(\textbf{y}_{i}\textbf{y}_{i}^H) = \left\{ \begin{array}{ll}
\textbf{F}_i(\sigma_s^2\textbf{H}_s\textbf{H}_s^H+\sigma_n^2\textbf{I})\textbf{F}_i^H, ~~~~~~~~~~~~~~ \textrm{for $i=1$,}\\\textbf{F}_i[\textbf{H}_{i-1,i}\textbf{A}_{i-1}E(\textbf{y}_{i-1}\textbf{y}_{i-1}^H)\textbf{A}_{i-1}^H\textbf{H}_{i-1,i}^H
+\sigma_n^2\textbf{I}]\textbf{F}_i^H\\~~~~~~~~~~~~~~~~~~~~~~~~~~~~~~~~\textrm{for $i=2,3,...,m$},\\
\end{array}\right.
\end{equation}

\begin{equation}
E(\textbf{y}_{i}\textbf{s}^H) = \left\{ \begin{array}{ll}
\sigma_s^2\textbf{F}_i\textbf{H}_s, & ~~~~~~~~~~\textrm{for $i=1$,}\\\textbf{F}_i\textbf{H}_{i-1,i}\textbf{A}_{i-1}E(\textbf{y}_{i-1}\textbf{s}^H), & \textrm{for $i=2,3,...,m$}.\\
\end{array}\right.
\end{equation}

\end{document}